\input{aipcheck}

\documentclass[
    ,final            
  ]
  {aipproc}

\usepackage{mathrsfs,amsmath,amssymb,amsfonts}
\usepackage{bm}
\usepackage{graphicx}
\usepackage{dcolumn}

\layoutstyle{8x11single}

\begin{document}

\title{The Nuclear Physics of Neutron Stars}


\author{J. Piekarewicz}
 {address={Department of Physics, Florida State University,
                  Tallahassee, FL 32306-4350, USA}}

\begin{abstract}
 We explore the unique and fascinating structure of neutron stars. Although neutron 
 stars are of interest in many areas of Physics, our aim is to provide an intellectual 
 bridge between Nuclear Physics and Astrophysics. We argue against the naive 
 perception of a neutron star as a uniform assembly of neutrons packed to enormous 
 densities. Rather, by focusing on the many exotic phases that are speculated to exist 
 in a neutron star, we show how the reality is different and far more interesting. 
\end{abstract}

\maketitle


\section{Historical Context}

Almost 30 years after the seminal work by Oppenheimer and Volkoff on the structure 
of neutron stars\,\cite{Opp39_PR55}, a young graduate student by the name of Jocelyn 
Bell detected a ``bit of scruff'' in the data arriving into her radio telescope. The arriving 
signal was ``pulsing'' with such an enormous regularity (once every 1.337\,302\,088\,331 
seconds) that she and her research advisor Anthony Hewish were convinced that they 
had detected a beacon from an extraterrestrial civilization. Initially dubbed as ``Little 
Green Man 1'', the source now known as radio pulsar ``PSR B1919+21'' was shortly 
identified as a rapidly rotating neutron star\,\cite{Hewish:1968}. 

The term ``neutron star'' appeared in writing for the first time in the 1933 proceedings 
of the the American Physical Society by Baade and Zwicky who wrote the now famous 
phrase: \emph{With all reserve we advance the view that supernovae represent 
the transitions from ordinary stars into ``neutron stars'', which in their final stages consist 
of extremely closed packed neutrons}\,\cite{Baade:1934}. Remarkably, this announcement 
came barely two years after the discovery of the neutron by Chadwick\,\cite{Chadwick:1932}. 
It appears, however, that the possible existence of dense stars that resemble ``one giant 
nucleus'' was already contemplated by Landau---even before the discovery of the neutron. 
For a detailed account of Landau's role in the history of neutron stars see 
Ref.\,\cite{Yakovlev:2012rd}.

Another towering figure in the history of neutron stars was Subrahmanyan Chandrasekhar 
(``Chandra''). As luck would have it, Chandra's fundamental discovery---the prediction of a 
maximum white-dwarf mass known as the ``Chandrasekhar 
limit''\,\cite{Chandrasekhar:1931}---came a year before the discovery of the neutron. Yet, 
fully aware of this fundamental discovery, Chandra wrote: \emph{For a star of small mass the 
white-dwarf stage is an initial step towards complete extinction. A star of large mass cannot 
pass into the white-dwarf stage and one is left speculating on other possibilities}. We now know 
that these ``other possibilities'' refer to either a neutron star or a black hole. For his contribution 
to the understanding of physical processes of importance to the structure and evolution of stars, 
Chandra was recognized with the 1983 Nobel Prize.

Unfortunately, Jocelyn Bell was never awarded the Nobel Prize, although her advisor Anthony 
Hewish was recognized with the award in 1974 for ``his decisive role in the discovery of pulsars''. 
The exclusion of Jocelyn Bell as co-recipient of the Nobel Prize was both controversial and 
roundly condemned by the astrophysics community. Still, Bell has always displayed enormous 
grace and humility in the face of this controversy. Indeed, she has stated: \emph{I believe 
it would demean Nobel Prizes if they were awarded to research students, except in very 
exceptional cases, and I do not believe this is one of them.}  It appears that Dr. Iosif 
Shklovsky, as well as many others, did not share her views. Dr. Shklovsky---recipient of the 
1972 Bruce Medal for outstanding lifetime contributions to astronomy---told Jocelyn
Bell: \emph{Miss Bell, you have made the greatest astronomical discovery of the twentieth 
century.}

\section{Introduction}
\label{Introduction}

A neutron star is a gold mine for the study of physical phenomena that cut across a variety 
of disciplines, such as particle physics, nuclear physics, astrophysics, and general relativity
among others. Although the most common perception of a neutron star is that of a uniform 
assembly of neutrons packed to densities that may exceed that of normal nuclei by up to 
an order of magnitude, the reality is far different and significantly more interesting. On the 
one hand, neutron stars are not expected to be particularly massive. Recent observations 
of two accurately measured neutron stars suggest masses near 2\,$M_{\odot}$. These
represent the most massive neutron stars observed to 
date\,\cite{Demorest:2010bx,Antoniadis:2013pzd}. Although more massive stars may very 
well exist, it is widely believed that the maximum mass of a neutron star will not exceed 
3 solar masses\,\cite{Lattimer:2006xb}. On the other hand, neutron stars are extremely 
compact. While enormous progress has been made in constraining stellar radii, a simple 
estimate derived from the observed spin period of rapidly rotating neutron stars is highly 
insightful. The first discovered millisecond pulsar PSR 1937+21 is 
a remarkable neutron star that rotates almost 640 times a second with a spin period of 
$P\!=\!1.557\,806\,448\,872$\,milliseconds. This rate is nearly thousand times faster than 
Bell's PSR B1919+21 pulsar. Given such an enormous spin frequency, the pulsar must be 
compact enough to allow the gravitational attraction to balance the immense centripetal
acceleration. This simple argument places an upper limit on the stellar radius of
\begin{equation}
 R \lesssim \sqrt[3]{r_{s}^{\odot} \left(\frac{cP}{2\pi}\right)^{2}
 \left(\frac{M}{2M_{\odot}}\right)} \approx 15\,{\rm km}\;,
\end{equation}
where $r_{s}^{\odot}\!\approx\!3$\,km is the Schwarzschild radius of the Sun, namely, the radius 
at which the Sun would become a black hole. Naturally, an object with such exotic properties displays 
unique characteristics. In Table\,\ref{Table1} we list some of the characteristics of a ``canonical'' 
$1.4\,M_{\odot}$ neutron star, such as the Crab pulsar---the compact remnant of the 1054 supernovae 
explosion that was witnessed and recorded in multiple Chinese and Japanese documents. 
\begin{center}
\begin{table}[h]
\begin{tabular}{l|l}
\hline
  Name: PSR B0531+21 & Constellation: Taurus\\
  Distance: 2.2 kpc  & Age: 960 years\\
  Mass: $1.4\,M_{\odot}$ & Radius: 10\,km \\ 
  Density: $10^{15}$g/cm${}^{3}$ & Pressure: $10^{29}$\,atm \\  
  Surface Temperature: $10^{6}$\,K & Escape velocity: 0.6\,c\\
  Period: 33\,ms & Magnetic Field: $10^{12}$ G\,\\
\hline
\end{tabular}
\caption{Approximate characteristics of a ``canonical'' neutron star, such as the 
960 year old Crab pulsar.}
\label{Table1}
\end{table}
\end{center}
\vspace{-0.5cm}

How does one describe the structure of such an exotic object. Spherically symmetric
neutron stars in hydrostatic equilibrium satisfy the Tolman-Oppenheimer-Volkoff (TOV) 
equations\,\cite{Opp39_PR55,Tol39_PR55}, which represent an extension of Newton's 
laws to the domain of general relativity. The TOV equations form a coupled set of first-order 
differential equations of the following form:
\begin{subequations}
 \begin{align}
   & \frac{dP}{dr}=-G\,\frac{{\cal E}(r)M(r)}{r^{2}}
         \left(1+\frac{P(r)}{{\cal E}(r)}\right)
         \left(1+\frac{4\pi r^{3}P(r)}{M(r)}\right)
         \left(1-\frac{2GM(r)}{r}\right)^{-1} \;,
         \label{TOVa}\\
   & \frac{dM}{dr}=4\pi r^{2}{\cal E}(r)\;,
         \label{TOVb}
 \end{align}
 \label{TOV}
\end{subequations}
where $P(r)$, ${\cal E}(r)$, and $M(r)$ represent the pressure, energy density, and (enclosed)
mass profiles of the star, and $G$ is Newton's gravitational constant. For a given fluid element 
in the star, hydrostatic equilibrium is attained by adjusting the pressure gradient to exactly 
balance the gravitational pull. Note that the last three terms in Eq.\,(\ref{TOVa}) (enclosed in 
parentheses) are of general-relativistic origin. For a neutron star with an escape velocity of about 
half of the speed of light, the effects from general relativity (GR) are essential. For example, 
ignoring GR effects would lead to a ``Chandrasekhar limit'' for a neutron star supported exclusively 
by neutron degeneracy pressure of $M_{\rm ch}\!\approx\!4\!\times\!1.4\,M_{\odot}\!=\!5.6\,M_{\odot}$. 
By properly incorporating GR corrections---which effectively enhance the pull from gravity---Oppenheimer 
and Volkoff determined a maximum neutron-star mass of only 
$M_{\rm max}\!\approx\!0.7\,M_{\odot}$\,\cite{Opp39_PR55}. Given that most accurately measured 
neutron-star masses fall in the $1.3\!-\!\!1.6\,M_{\odot}$ range\,\cite{Lattimer:2012nd}, it is interesting 
to quote from Oppenheimer and Volkoff: ``It seems likely that our limit of $\sim\!0.7M_{\odot}$ is near 
the truth''\,\cite{Opp39_PR55}. Although there is nothing wrong with the calculation by Oppenheimer 
and Volkoff, it does suffer from a critical omission: the role of nuclear interactions in modifying the 
equation of state (EOS) of a free Fermi gas of neutrons. Indeed, at the enormous densities encountered 
in a neutron star, the strong short-repulsion of the nucleon-nucleon (NN) force can not be ignored. It is 
precisely the critical role of nuclear interactions that makes neutron stars such a fruitful ground for the 
study of nuclear physics at the extremes of densities and isospin asymmetry. Remarkably, the only 
input that neutron stars are sensitive to is the equation of state of neutron-rich matter, namely, a 
relation between the pressure and the energy density $P\!=\!P({\cal E})$. Conversely and equally 
remarkable, each EOS generates a \emph{unique} mass-{vs}-radius relation\,\cite{Lindblom:1992}.

In hydrostatic equilibrium the neutron star is perfectly balanced by the action of two enormous forces: 
gravity and pressure. As indicated in Eq.\,(\ref{TOVa}), hydrostatic equilibrium demands that the 
pressure gradient $dP/dr$ be negative, so the pressure decreases monotonically with distance until it 
vanishes at the edge of the star. In particular, the highest pressure---and density---at the center of the 
star must be enormous in order to be able to support the full weight of the star. This requires densities 
that may significantly exceed the one at the center of the nucleus 
(i.e., $\rho_{{}_{\!0}}\!=\!2.48\times 10^{14}{\rm g/cm^{3}}$). This implies that models of the EOS will 
have to encompass density regions---both high and low---inaccessible to experiment. What novel phases 
emerge under such extreme conditions is both fascinating and unknown. In what follows we embark on a 
journey of a neutron star that highlights such fascinating phases and discusses the tools that are required 
to uncover their observable signatures.

\section{A Journey through a Neutron Star}
\label{Tour}

\begin{figure}[h]
  \includegraphics[width=0.35\columnwidth,height=6cm]{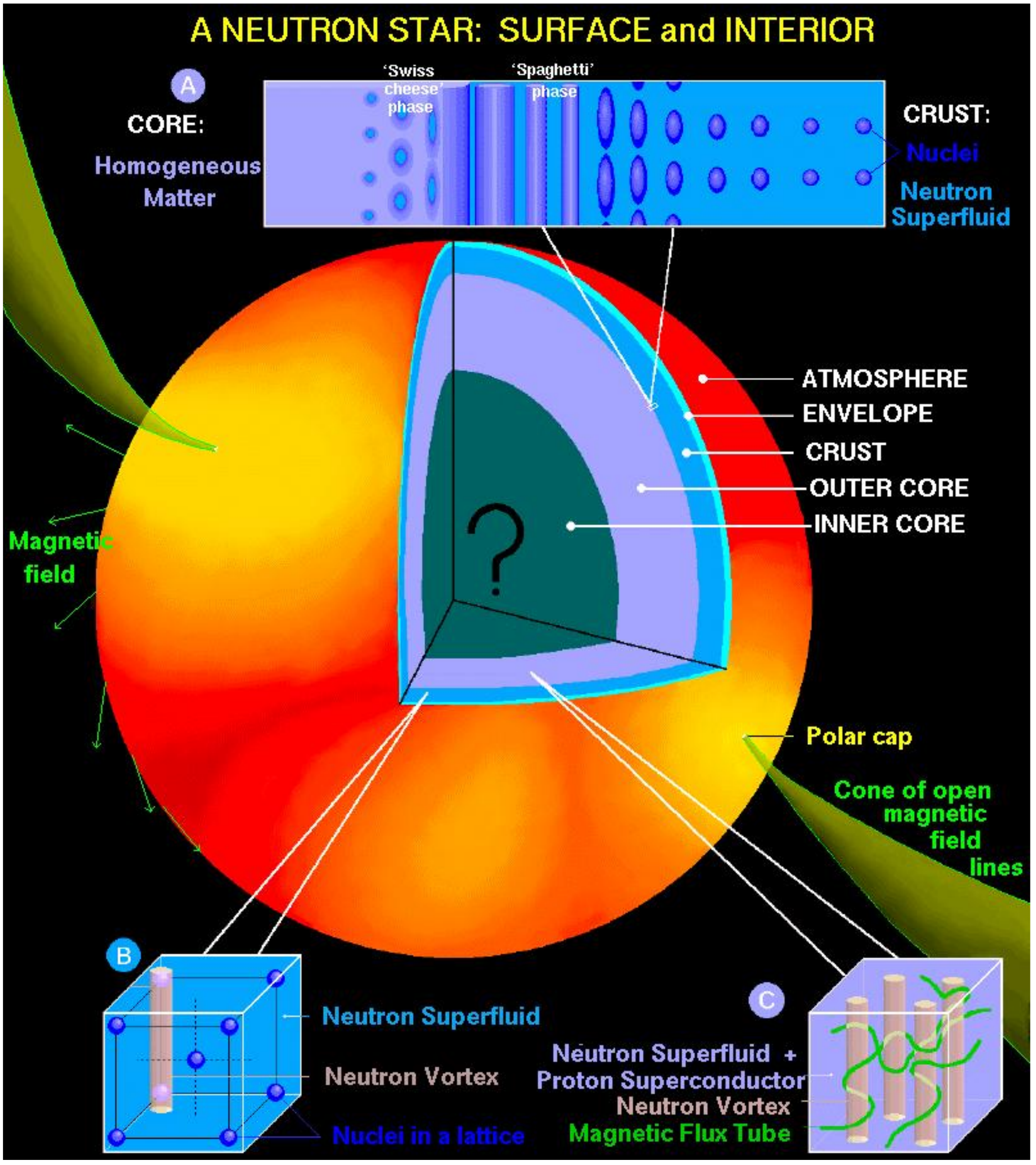}
    \hspace{0.75cm}
  \includegraphics[width=0.40\columnwidth,height=6cm]{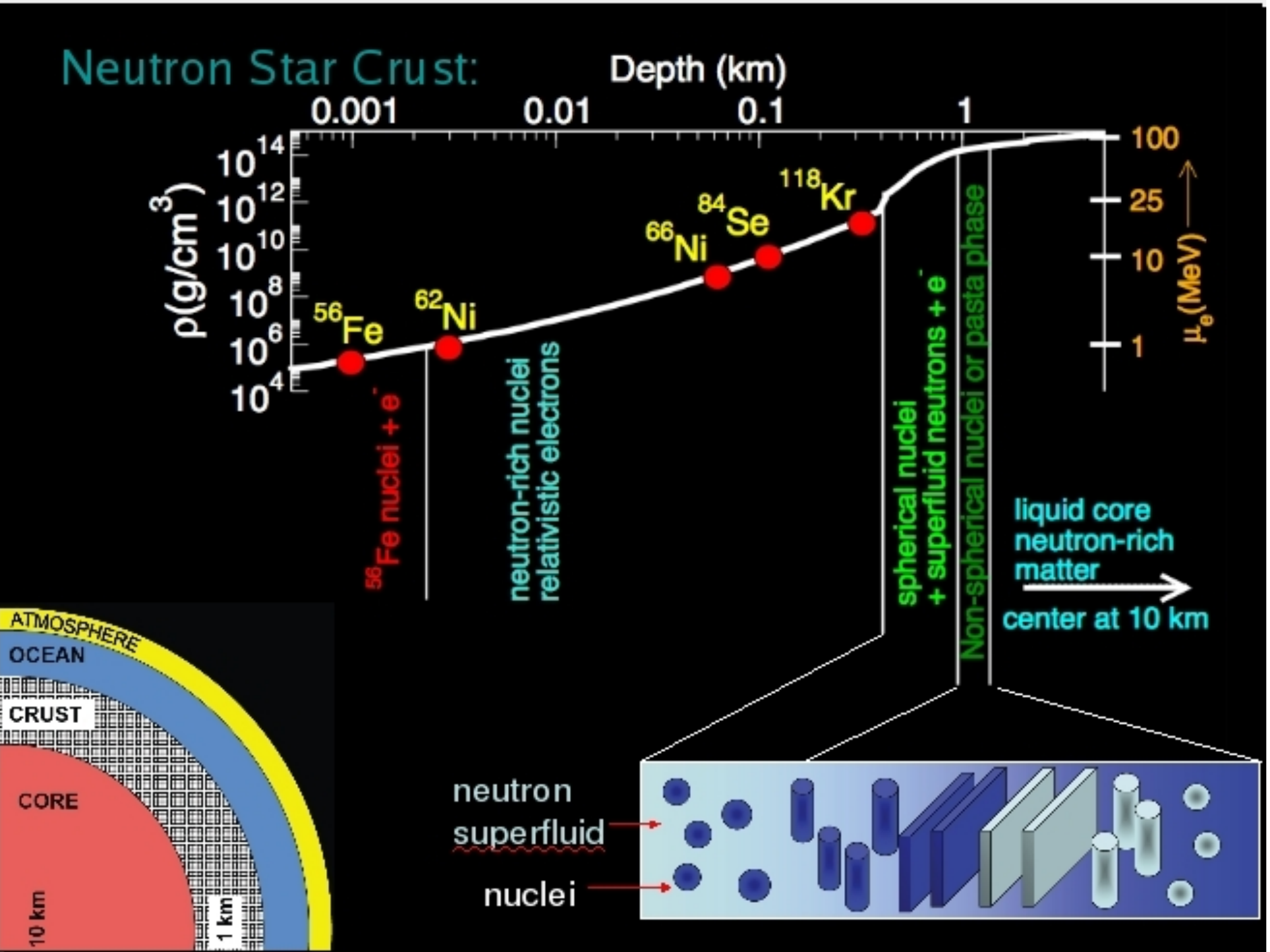}
   \caption{(Left panel) An accurate rendition of the structure and phases of a neutron star---courtesy of Dany Page.
             (Right panel) An accurate depiction of the composition of the crust of a neutron star---courtesy of Sanjay Reddy.}
 \label{Fig1}
\end{figure}

The structure of neutron stars is both interesting and complex. Fig.\,\ref{Fig1}a provides what is believed 
to be an accurate rendition of the structure of a neutron star. The outermost surface of the neutron star
contains a very thin atmosphere of only a few centimeters thick that is composed of Hydrogen, but may
also contain heavier elements such as Helium and Carbon. The electromagnetic radiation that we 
observe may be used to constrain critical parameters of the neutron star. For example, assuming 
blackbody emission from the stellar surface at a temperature $T$ provides a determination of the
stellar radius from the Stefan-Boltzmann law: $L\!=\!4\pi\sigma R^{2}T^{4}$. Unfortunately, complications
associated with distance measurements and distortions of the black-body spectrum make the accurate
determination of stellar radii a challenging task. Just below the atmosphere lies the $\sim\!100$\,m
thick envelope that acts as ``blanket'' between the hot interior (with $T\!\gtrsim\!10^{8}$\,K) and the
``cold'' surface (with $T\!\gtrsim\!10^{6}$\,K)\,\cite{Page:2004fy}. Further below lies the non-uniform crust, 
a 1\,km region that is believed to be involved in pulsar glitches---the sudden increase 
in the rotational frequency of the star. The non-uniform crust sits above a uniform liquid core that 
consists of neutrons, protons, electrons, and muons. The core accounts for practically all the mass 
and for about 90\% of the size of a neutron star. There is also a possibility, marked with a question mark 
in Fig.\ref{Fig1}a, of an inner core made of exotic particles, such as hyperons, meson condensates, and 
quark matter\,\cite{Page:2006ud}. Given that at this time there is no clear observational evidence of their 
existence, such possibility will no longer be addressed here. Instead, we now continue with a detailed 
discussion of the structure of the crust and the \emph{non-exotic} core.

\subsection{The Outer Crust}
\label{OuterCrust}

The outer crust is traditionally associated with the region of the star spanning about seven orders of magnitude 
in density; from about $10^{4}{\rm g/cm^{3}}$ to $4\!\times\!10^{11}{\rm g/cm^{3}}$~\cite{Baym:1971pw}. At 
these densities, the electrons---which form an essential component of the star in order to maintain charge 
neutrality---have been pressure ionized and move freely throughout the crust as a free Fermi gas. The dynamics 
of the outer crust is relatively simple, as it depends almost exclusively on nuclear masses. At these low 
sub-saturation densities, it is energetically favorable for nucleons to cluster into ${}^{56}$Fe nuclei that arrange 
themselves in a crystalline lattice. However, as the density increases ${}^{56}$Fe ceases to be the preferred nucleus. 
This is because the electronic contribution to the total energy increases rapidly with density. Thus, it becomes 
energetically advantageous for the energetic electrons to capture on protons and for the excess energy to be 
carried away by neutrinos. Thus at densities of about $10^{6}{\rm g/cm^{3}}$, ${}^{62}$Ni becomes the most stable 
nucleus. As the density continues to increase, the nuclear system evolves into a Coulomb lattice of progressively 
more exotic, neutron-rich nuclei\,\cite{RocaMaza:2008ja}; see Fig.\ref{Fig1}b for an accurate depiction of the 
composition of the crust. Finally, at a ``critical'' density of about $4\!\times\!10^{11}{\rm g/cm^{3}}$, nuclei are 
unable to hold any more neutrons and the neutron drip line is reached. Most mass models used in the literature 
predict that the sequence of progressively more exotic nuclei will terminate with ${}^{118}$Kr---a nucleus with 36 
protons and 82 neutrons. Note that the last isotope with a well measured mass is ${}^{97}$Kr, which is still 21 
neutrons away from ${}^{118}$Kr! Thus, one most rely on mass models that are often hindered by uncontrolled 
extrapolations. In this regard, mass measurement on exotic nuclei are critical. A recent landmark experiment 
at \emph{ISOLTRAP} measured for the first time the mass of ${}^{82}$Zn\,\cite{Wolf:2013ge}. The addition of 
this one mass value alone resulted in an interesting modification in the composition profile of the outer crust; 
see also Ref.\,\cite{Pearson:2011zz}.

\subsection{The Inner Crust}
\label{Inner Crust}

\begin{figure}[h]
  \includegraphics[width=0.45\columnwidth]{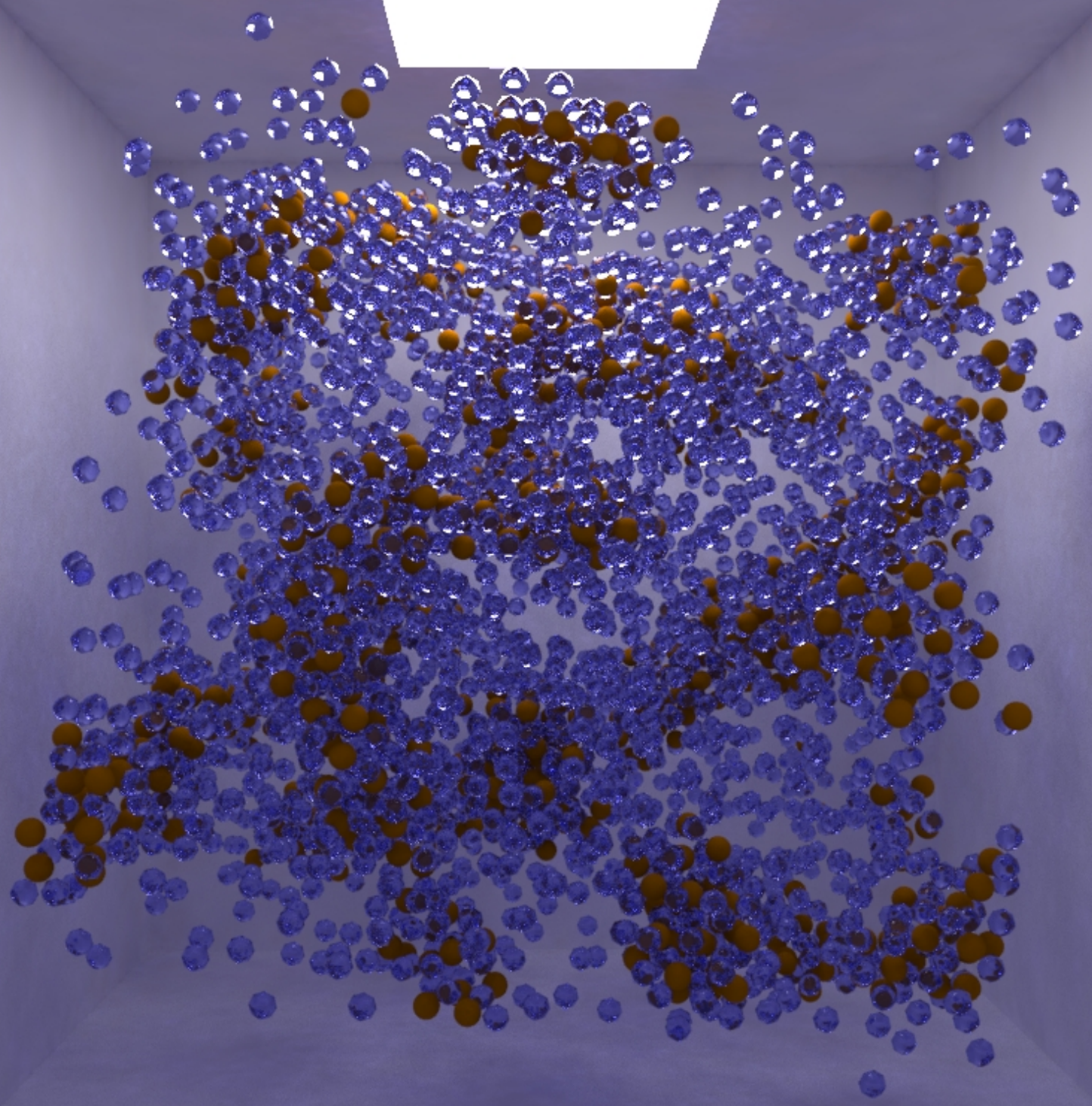}
 \caption{A snapshot of a Monte-Carlo simulation for a system consisting of 4000 
 nucleons at a baryon density of  $\rho\!=\!\rho_{{}_{\!0}}/6$, a proton fraction of $Z/A\!=\!0.2$, and 
 an effective temperature of $T\!=\!1$\,\,MeV\,\cite{Horowitz:2004yf,Horowitz:2004pv}.} 
 \label{Fig2}
\end{figure}

The inner stellar crust comprises the region from neutron-drip density up to the density at which uniformity in 
the system is restored; about a third of nuclear matter saturation density $\rho_{{}_{\!0}}$. On the top layers of 
the inner crust nucleons continue (as in the outer crust) to cluster into a Coulomb crystal of neutron-rich nuclei 
embedded in a uniform electron gas. Now, however, the crystal is in chemical equilibrium with a superfluid 
neutron vapor. Note that although the precise details of the pulsar glitch mechanism are 
unclear\,\cite{Andersson:2012iu}, the common perception is that glitches develop as a result of the tension 
created by the differential rotation between the superfluid component and the normal component, which continuously 
spins down due to the emission of electromagnetic radiation. As the density increases, the spherical nuclei that form 
the crystal lattice start to deform in an effort to reduce the Coulomb repulsion. As a result, the system starts to exhibit 
rich and complex structures that emerge from a dynamical competition between the short-range nuclear attraction 
and the long-range Coulomb repulsion. At the long length scales characteristic of the outer crust the system 
organizes itself into a crystalline lattice of well-separated spherical nuclei. At the other extreme of densities, 
uniformity in the core gets restored and the system behaves as a uniform Fermi liquid of 
nucleons and leptons. However, the transition region from the highly ordered crystal to the uniform liquid 
core is complex and poorly understood. Length scales that were well separated in both the crystalline and uniform 
phases are now comparable, giving rise to a universal phenomenon known as \emph{Coulomb frustration}. In the
bottom layers of the crust Coulomb frustration is manifested by the emergence of complex structures of various
topologies dubbed  ``nuclear pasta''\,\cite{Ravenhall:1983uh,Hashimoto:1984}. For some recent reviews on the 
fascinating structure and dynamics of the neutron-star crust see Refs.\,\cite{Chamel:2008ca,Bertulani:2012}, and 
references contain therein. In particular, in Fig.\,\ref{Fig2} we display a snapshot obtained from a numerical 
simulation of a system of $Z\!=\!800$\,protons and $N\!=\!3200$\,neutrons that illustrates how the system 
organizes itself into neutron-rich clusters of complex topologies that are immersed in a dilute neutron 
vapor\,\cite{Horowitz:2004yf,Horowitz:2004pv}. We note that a great virtue of the numerical simulations carried 
out in Refs.\,\cite{Horowitz:2004yf,Horowitz:2004pv} is that pasta formation is studied in an unbiased way 
without assuming any particular set of shapes. Rather, the system evolves dynamically into these 
complex shapes from an underlying two-body interaction consisting of a short-range nuclear attraction and a 
long-range Coulomb repulsion.
Although the dynamics of the pasta is indeed fascinating, its existence has never been verified either
experimentally or observationally. In the laboratory, heavy-ion collisions come closest at reproducing 
the conditions required for pasta formation. For example, low-energy heavy-ion collisions produce 
dilute neutron-rich matter that may closely resemble the conditions found in the bottom layers of the
inner crust. However, the matter produced in these collisions is ``warm'' and theoretical models may 
be required to extrapolate to the fully catalyzed regime found in neutron stars. However, an intriguing 
fact may provide the first observational manifestation of the pasta phase in neutron stars. It appears 
that the spin period in a certain population of neutron stars known as \emph{isolated X-ray pulsars} is 
constrained to at most 12 seconds long. That is, although magnetic fields estimates suggest that 
X-ray pulsars could slow down to periods of about 30 seconds in a few thousand years, none is observed 
with periods longer than 12 seconds\,\cite{Pons:2013nea}. Pons, Vigan\`o, and Rea have shown that the 
existence of a highly resistive layer in the inner crust---\emph{likely the pasta phase}---decreases the 
electrical conductivity, thereby resulting in a quenching of the dipolar magnetic field that ultimately limits 
the spin period to a maximum of about 20 seconds\,\cite{Pons:2013nea}.

\subsection{The Stellar Core}
\label{Core}

\begin{figure}[h]
 \includegraphics[width=0.45\columnwidth]{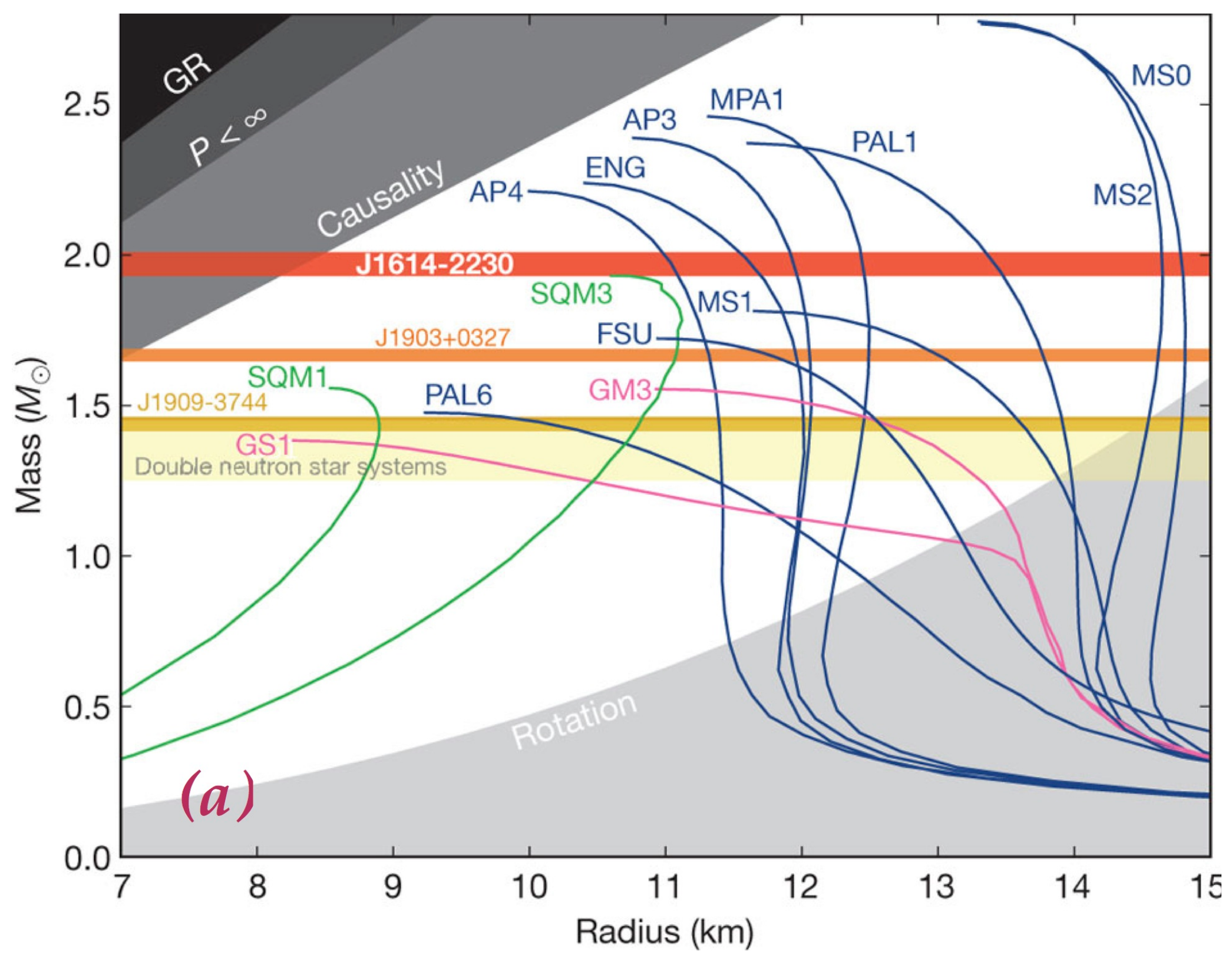}
  \hspace{0.75cm}
 \includegraphics[width=0.48\columnwidth]{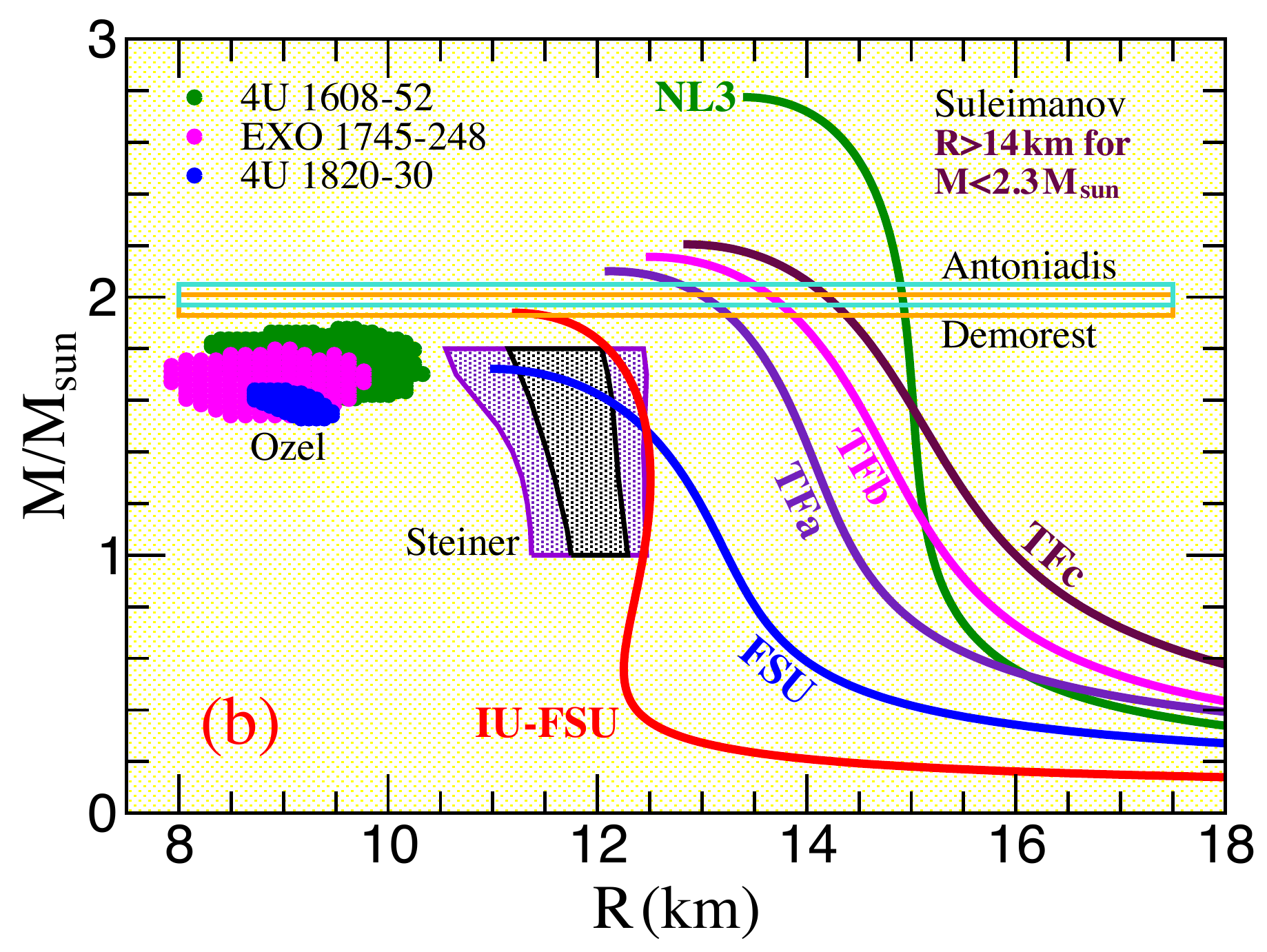}
 \vspace{-0.2cm}
   \caption{(Left panel) Predictions for the mass-vs-radius relation for a variety of models of the 
                EOS using both exotic and non-exotic cores\,\cite{Demorest:2010bx}. (Right panel) 
                Constraints on both stellar masses and radii extracted from various analyses of X-ray 
                bursts\,\cite{Ozel:2010fw,Steiner:2010fz,Suleimanov:2010th}. Also shown are 
                constraints obtained from the measurement of two massive neutron stars 
                by Demorest\,\cite{Demorest:2010bx} and Antoniadis\,\cite{Antoniadis:2013pzd}.}
 \label{Fig3}
\end{figure}

At densities of about $10^{14}{\rm g/cm^{3}}$ the pasta phase will ``melt'' and uniformity in the system will be 
restored.  At these densities (of about a third to a half of $\rho_{{}_{\!0}}$) the original perception of Baade and 
Zwicky\,\cite{Baade:1934}, namely, that of a neutron star as a uniform assembly of extremely closed packed 
neutrons, is finally realized in the stellar core. Note, however, that in order to maintain both chemical equilibrium 
and charge neutrality a small fraction of about 10\% of protons and leptons is required. Whereas the crust displays 
fascinating and intriguing dynamics, its structural impact on the star is rather modest. Indeed, more than 90\% of
the size and most of the mass reside in the stellar core. However, the equation of state of neutron-rich matter at 
the high densities attained in the core is poorly constrained by laboratory observables. Perhaps the cleanest
constraint on the EOS at high-density will emerge as we answer one of the fundamental questions in nuclear 
astrophysics: \emph{what is the maximum mass of a neutron star?} Or equivalently, \emph{what is the 
minimum mass of a black hole?} In this regard, enormous progress has been made with the recent observation 
of two massive neutron stars by Demorest et al.\,\cite{Demorest:2010bx} and Antoniadis et al.\,\cite{Antoniadis:2013pzd}. 
Figure\,\ref{Fig3}a displays the major impact of the mass measurement of PSR J164-2230 
$(1.97\!\pm\!0.04\,M_{\odot})$ as this measurement alone can rule out EOS that are too soft to support a 
$2\,M_{\odot}$ neutron star---such as those with exotic cores. Undoubtedly, the quest for even more massive 
neutron stars will continue with the commissioning of the \emph{Large Observatory for X-ray Timing} (LOFT)
that will provide valuable insights into the behavior of ultra-dense matter.
In contrast, the accurate determination of stellar radii has proven more challenging. However, observations 
of a variety of spectroscopic phenomena in X-ray bursters appear to provide a promising approach for resolving 
both the mass and radius of a neutron star. Results from the analysis of three X-ray bursters by \"Ozel, Baym, 
and G\"uver\,\cite{Ozel:2010fw} (shown in Fig.\,\ref{Fig3}b) suggest very small radii---of $8$ to 
$\!10$\,km---that are difficult to reconcile with the predictions from models lacking exotic cores\,\cite{Fattoyev:2010rx}. 
Indeed, \emph{none of the models} displayed in Fig.\,\ref{Fig3} can account for such a small stellar radius and maximum 
masses of at least $2\,M_{\odot}$. However, it has been recognized that systematic uncertainties in the analysis 
of X-ray bursters, first by Steiner, Lattimer, and Brown\,\cite{Steiner:2010fz} and shortly after by Suleimanov 
et al.\,\cite{Suleimanov:2010th}, may invalidate the results of Ref.\,\cite{Ozel:2010fw}.  
The results by Steiner et al.\,\cite{Steiner:2010fz}, depicted in Fig.\,\ref{Fig3}b by the two shaded areas that 
indicate 1$\sigma$ and 2$\sigma$ contours, suggest stellar radii in the $10$-$\!13$ km range. However,
to complicate things even further, Suleimanov et al. have proposed a \emph{lower limit} on the stellar radius of 
14\,km for neutron stars with masses below 2.3\,M$_{\odot}$, concluding that neutron-star matter is characterized 
by a stiff EOS\,\cite{Suleimanov:2010th}. One must then conclude that whereas thermal emissions during X-ray 
bursts may become a powerful tool in the determination of stellar radii, at present such analyses are plagued 
by numerous uncertainties.

However, the situation seems to have improved recently with the study of accreting
neutron stars during quiescence, the so-called \emph{quiescent low-mass X-ray binaries} (qLMXBs). By studying 
the thermal spectra of five qLMXBs inside globular clusters, Guillot and collaborators reported a \emph{common 
radius} for all five sources of $R_{NS}\!=\!9.1^{+1.3}_{-1.5}$\,km at a 90\% confidence level\,\cite{Guillot:2013wu}. 
Such small radius seriously challenges our current understanding of the equation of state of dense matter. 
Indeed, it appears that few realistic models (if at all!) can accommodate both a small stellar radius and a large
limiting mass\,\cite{Wiringa:1988tp}. One should mention that whereas the approach developed in 
Ref.\,\cite{Guillot:2013wu} provides a careful accounting of all uncertainties, some of the assumptions---such as 
a single common radius---and some of the adopted uncertainties have been called into 
question\,\cite{Lattimer:2013hma}. However, we are confident that with the launching of space missions such 
as LOFT and GAIA---with the latter providing \emph{unprecedented positional measurements for about one billion 
stars in our Galaxy}---many of the current problems will be mitigated.

\begin{figure}[h]
 \includegraphics[width=0.5\columnwidth]{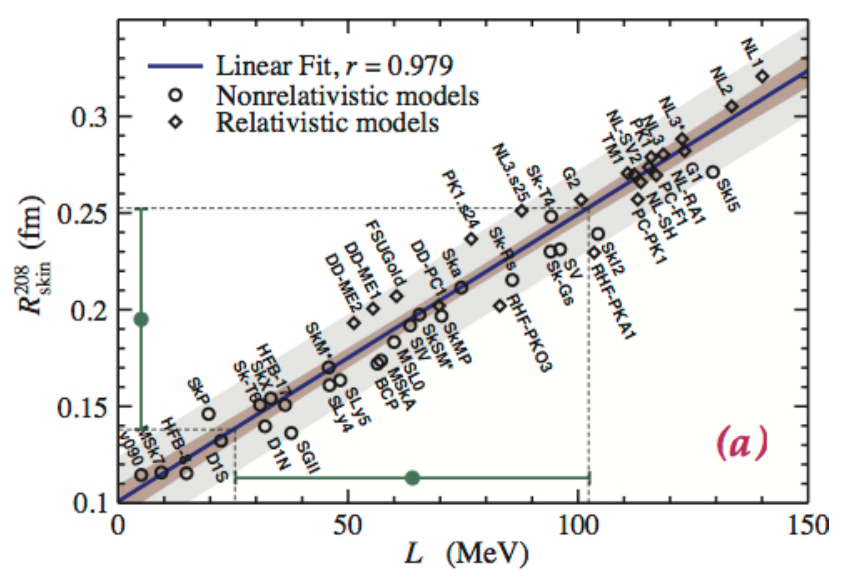}
  \hspace{0.75cm}
 \includegraphics[width=0.35\columnwidth]{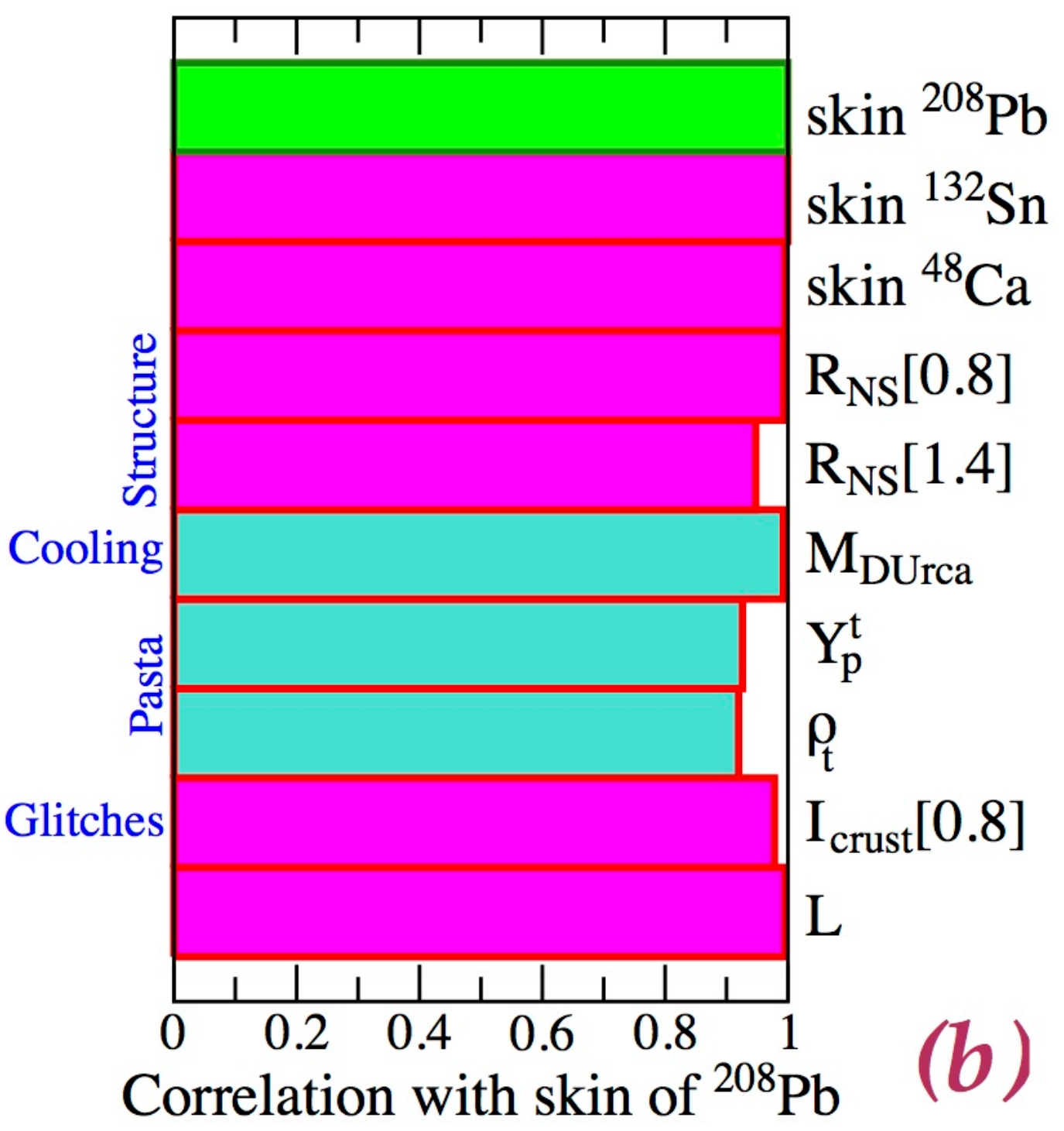}
 \vspace{-0.2cm}
  \caption{(Left panel) Predictions from a large number of nuclear density functionals for the 
  	       neutron-skin thickness of ${}^{208}$Pb and the slope of the symmetry
	       energy $L$\,\cite{RocaMaza:2011pm}. Constraints from an updated PREX measurement
	       (``PREX-II'') have been incorporated into the plot. (Right panel) Correlation
	       coefficients between $R_{\rm skin}^{208}$ and a variety of neutron-star properties 
	       as predicted by the FSUGold density functional\,\cite{Todd-Rutel:2005fa}.} 
 \label{Fig4}
\end{figure}

Whereas laboratory experiments are of marginal utility in constraining the limiting mass of a neutron star, they
play an essential role in constraining stellar radii. This is because the radius of a neutron star is sensitive to 
the density dependence of the \emph{symmetry energy} in the immediate vicinity of nuclear-matter saturation 
density\,\cite{Lattimer:2006xb}. Note that the symmetry energy represents the energy cost required to convert 
symmetric nuclear matter into pure nuclear matter. A critical property of the EOS that has received considerable 
attention over the last decade is the slope of the symmetry energy at saturation density\,\cite{Piekarewicz:2008nh}. 
The slope of the symmetry energy $L$ is directly related to the pressure of pure neutron matter at $\rho_{{}_{\!0}}$
and, as such, is strongly correlated to a myriad of neutron-star observables, such as its composition and cooling 
dynamics\,\cite{Fattoyev:2012rm}. Moreover, $L$ is also strongly correlated to the thickness of the neutron skin 
of heavy nuclei\,\cite{Brown:2000,Furnstahl:2001un}---which is defined as the difference between the neutron and 
proton root-mean-square radii. The physical reason behind this correlation is particularly insightful. Heavy nuclei 
contain an excess of neutrons as a result of the repulsive Coulomb interaction among the protons. Energetically, 
it is advantageous---to both the surface tension and to the symmetry energy---to form an isospin symmetric ($N\!=\!Z$)
core. So the basic question is \emph{where do the extra neutrons go?} Placing them in the core reduces the surface
tension but increases the symmetry energy. In contrast, moving them to the surface increases the surface tension
but reduces the symmetry energy---which is lower in the dilute surface than in the dense core. So the thickness of
the neutron skin emerges from a competition between the surface tension and the \emph{difference} between the 
value of the symmetry energy at the surface relative to that at the center (i.e., $L$). If such a difference is large,
then it is favorable to move many neutrons to the surface, thereby creating a thick neutron skin. This suggests
a powerful correlation: \emph{the larger the value of $L$ the thicker the neutron skin}\,\cite{Horowitz:2001ya}.
Indeed, the strong correlation between $L$ and the neutron-skin thickness of ${}^{208}$Pb ($R_{\rm skin}^{208}$)
is displayed in Fig.\,\ref{Fig4}a where a large and representative set of density functionals were used to predict 
both\,\cite{RocaMaza:2011pm}. The strong correlation coefficient of $r\!=\!0.979$ suggests how a laboratory observable 
such as $R_{\rm skin}^{208}$ may serve to determine a fundamental property of the equation of state.

Recently, the Lead Radius Experiment (``PREX'') at the Jefferson Laboratory has provided the first model-independent 
evidence in favor of a neutron-rich skin in ${}^{208}$Pb\,\cite{Abrahamyan:2012gp,Horowitz:2012tj}. Given that the 
neutral weak vector boson couples strongly to the neutron, parity violating electron scattering provides a clean probe of 
neutron densities that is free from strong-interaction uncertainties. As the proton radius of ${}^{208}$Pb is known extremely 
accurately from conventional (parity conserving) electron scattering, PREX effectively determined the neutron-skin 
thickness of  ${}^{208}$Pb to be: $R_{\rm skin}^{208}\!=\!{0.33}^{+0.16}_{-0.18}\,{\rm fm}$\,\cite{Abrahamyan:2012gp}. 
In the future, an updated and approved PREX measurement---``PREX-II''---will improve the determination of $R_{\rm skin}^{208}$ 
by a factor of 3 (i.e., an error of about 0.06\,fm is anticipated). The impact of such a measurement in constraining the symmetry 
pressure $L$ is displayed with green error bars in Fig.\,\ref{Fig4}a. Remarkably, the same pressure that is responsible for 
creating a neutron-rich skin supports a neutron star against gravitational collapse. Thus, models that predict thicker neutron 
skins often produce neutron stars with larger radii\,\cite{Horowitz:2000xj,Horowitz:2001ya}. This makes possible to establish 
``data-to-data'' relation between the neutron-rich skin of a heavy nucleus and the radius of a neutron star. This fact is nicely 
illustrated in Fig.\,\ref{Fig4}b that shows the enormous reach of an accurate measurement of $R_{\rm skin}^{208}$. 
The correlations displayed in the figure were investigated through a covariance analysis\,\cite{Fattoyev:2012rm} using 
the accurately-calibrated FSUGold density functional\,\cite{Todd-Rutel:2005fa}. Besides the expected correlation between 
$R_{\rm skin}^{208}$  and the neutron-skin thickness of other neutron-rich nuclei, the figure displays a strong correlation 
with a variety of neutron-star observables---including neutron-star radii. For a detailed explanation of the physics behind this 
plot see Ref.\,\cite{Fattoyev:2012rm}.

\section{Conclusions}
\label{Conclusions}

Neutron stars provide a powerful intellectual bridge between Nuclear Physics and Astrophysics. In this manuscript 
we explored the fascinating structure of neutron stars and discussed how critical laboratory experiments and 
astronomical observations may constrained the EOS. In the particular case of the outer crust, we established the 
fundamental role that mass measurements of exotic nuclei at rare isotope facilities will play in elucidating its distinct 
composition. Next, we moved to the deeper inner crust with its complex pasta phase shapes that emerge from 
Coulomb frustration and which display unique dynamical features. Although finding clear signatures of its existence 
has proved elusive, we discussed how the lack of X-ray pulsars with long spin periods may be the first observable 
manifestation of the nuclear pasta phase. Finally, we moved into the deep stellar core and reported on a recent 
analysis of quiescent low-mass X-ray binaries that argues in favor of small neutron-star radii. This finding---suggesting 
a soft EOS---poses serious challenges to theoretical models that must at the same time account for the existence of 
massive neutron stars---which instead suggests a stiff EOS. Moreover, small stellar radii may also be at odds with the 
PREX finding of a rather large neutron-skin thickness in ${}^{208}$Pb (although the errors are large). If future laboratory 
experiments and astronomical observations confirm that both $R_{\rm skin}^{208}$ is thick and stellar radii are small, this 
would strongly suggest a softening of the EOS due to the onset of a phase transition. However, the EOS must eventually 
stiffen to account for the existence of massive neutron stars. Such extraordinary behavior will confirm the unique role of 
neutrons stars as gold mines for the study of the EOS of dense nucleonic matter.

\begin{theacknowledgments}
  The author thanks the organizers of the Seventh European Summer School on Experimental Nuclear Astrophysics
  for their kind hospitality. This work was supported in part by grant DE-FD05-92ER40750 from the Department of 
  Energy.
\end{theacknowledgments}

\bibliographystyle{aipproc}   
\bibliography{../../ReferencesJP}
\end{document}